\documentclass{ws-ijmpd}

\def\AJ{{\it Astroph. J.} }

\def\CQG{{\it Class. Quantum Gravity} }

\def\GRG{{\it Gen. Relat. Gravit.} }

\def\IJMP{{\it Int. J. Mod. Phys.} }

\def\MNRAS{{\it Mon. Not. R. Ast. Soc.} }
\def\NAT{{\it Nature} }

\def\PL{{\it Phys. Lett.} }
\def\PR{{\it Phys. Rev.} }
\def\PRL{{\it Phys. Rev. Lett.} }

\def\frac#1#2{{\textstyle{{#1}\over {#2}}}}

\def\lsim{\mathrel{\rlap{\lower4pt\hbox{\hskip1pt$\sim$}}
    \raise1pt\hbox{$<$}}}
\def\gsim{\mathrel{\rlap{\lower4pt\hbox{\hskip1pt$\sim$}}
    \raise1pt\hbox{$>$}}}
\def\sqr#1#2{{\vcenter{\vbox{\hrule height.#2pt
         \hbox{\vrule width.#2pt height#1pt \kern#1pt
         \vrule width.#2pt}
         \hrule height.#2pt}}}}

\def\beq{\begin{equation}}
\def\eeq{\end{equation}}
\def\beqa{\begin{eqnarray}} 
\def\eeqa{\end{eqnarray}}

\def\laq{\raise 0.4 ex \hbox{$<$}\kern -0.8 em\lower 0.62 ex\hbox{$\sim$}}
\def\gaq{\raise 0.4 ex \hbox{$>$}\kern -0.7 em\lower 0.62 ex\hbox{$\sim$}}

\begin{document}

\markboth{Orfeu Bertolami}
{Dark Energy, Dark Matter and Gravity}

%
\catchline{}{}{}{}{}
%

\title{Dark Energy, Dark Matter and Gravity\footnote{Talk presented at the International 
Workshop ``From Quantum to Cosmos: Fundamental Physics Research in Space'', 22-24 May 2006, Warrenton, Virginia, USA.}}

\author{Orfeu Bertolami\footnote{E-mail: orfeu@cosmos.ist.utl.pt}}

\address{Instituto Superior T\'ecnico, Departamento de F\'\i sica \\
Av. Rovisco Pais 1, Lisbon, 1049-001, Portugal}




\maketitle


\begin{abstract}
We discuss the 
motivation for high accuracy relativistic gravitational experiments 
in the Solar System and complementary cosmological tests. 
We focus our attention on the issue of distinguishing a generic 
scalar-theory of gravity as the underlying 
physical theory from the usual general relativistic picture, where one 
expects the presence of fundamental scalar fields associated, for instance, 
to inflation, dark matter and dark energy.

\keywords{Dark matter; dark energy; scalar fields; gravity.}
\end{abstract}

\vspace{0.5cm}

\section{Introduction}
\label{sec:intro}

Present day experimental evidence indicates that  
gravitational physics is in agreement with Einstein's theory of
General Relativity to considerable accuracy; however, there are a number 
of reasons, theoretical and experimental, to question the theory as the 
ultimate description of gravity.

On the theoretical side, difficulties arise from various corners, most 
stemming from the strong gravitational field
regime, associated with the existence of spacetime singularities and the
difficulty to describe the physics of very strong gravitational
fields. Quantization of gravity is a possible way to overcome these 
obstacles, however, despite the success of modern gauge field theories in describing the
electromagnetic, weak, and strong interactions, the path to 
describe gravity at the quantum level is still to be found.
Indeed, our two foundational theories, Quantum Mechanics and
General Relativity, are not compatible with each other. Furthermore, in 
fundamental
theories that attempt to include gravity, new long-range forces
often arise in addition to the Newtonian inverse-square law. Even at
the classical level, and assuming the validity of the
Equivalence Principle, Einstein's theory does not provide the most
general way to establish the spacetime metric. There are
also important reasons to consider additional fields, especially
scalar fields. Although the latter appear in unification theories,
their inclusion predicts a non-Einsteinian
behavior of gravitating systems. These deviations from General
Relativity include violations of the Equivalence Principle, modification 
of large-scale gravitational phenomena, and variation of the
fundamental ``constants''. These predictions motivate new searches
for very small deviations of relativistic gravity from General
Relativity and drive the need for further gravitational experiments in space. 
These include 
laser astrometric measurements\cite{solvang_lator04,Lator01,Texas@Stanford_lator,ESTEC_lator}, 
high-resolution lunar laser ranging (LLR)\cite{Murphy_etal_2002} and 
long range tracking of spacecraft using the formation flight 
concept, as proposed\cite{Pioneer} to test 
the Pioneer anomaly\cite{Anderson02}. A broader discussion on the motivations to perform fundamental physics 
experiments in space can be found elsewhere\cite{Matos04}.

On the experimental front, recent cosmological
observations does lead one to conclude that our 
understanding of the origin and evolution of the
Universe based on General Relativity requires that most of the energy content of the Universe
resides in the presently unknown dark matter and dark energy 
components that may
permeate much, if not all spacetime. 
Indeed, recent Cosmic Microwave Background Radiation (CMBR) WMAP three year data\cite{WMAP3} 
indicates that our Universe is well described, 
within the framework of General Relativity, by a flat Robertson-Walker 
metric, meaning that the energy density of the Universe is fairly close to the critical one, 
$\rho_c \equiv 3H_0^2/8 \pi G \simeq 10^{-29} g/cm^3$, 
where $H_0 \simeq 73~km~s^{-1}Mpc^{-1}$ is the Hubble expansion parameter at present. 
Moreover, CMBR, 
Supernova and large scale structure data are consistent with 
each other if, in the cosmic budget of energy, dark energy corresponds to about $73\%$ of the critical density, 
while dark matter to about $23\%$ and baryonic matter, the matter that we are made of, to only about $4\%$.  
Furthermore, it is generally believed  
that the ultimate theory that will reconcile
Quantum Mechanics and General Relativity will also allow for 
addressing the cosmological questions related with the origin and destiny of the Universe. 

It is our opinion that the crystallization of these fundamental 
questions is well timed with recent progress 
in high-precision 
measurement technologies for physics experiments in space. This puts us
in position to realistically
address crucial questions, such as the nature of dark energy and dark matter,
the existence of intermediate range forces and the ultimate nature 
of gravity. Furthermore, given the ever
increasing practical significance of General Relativity, for 
spacecraft navigation, time transfer,
clock synchronization, weight and length standards, it is just 
natural to expect that the 
theory will be regularly tested with ever increasing accuracy.
Thus, it seems legitimate to speculate that the present state of 
physics represents a unique confluence 
of important challenges in high energy physics and cosmology together with 
technological advances and access to space, a   
conjunction that is likely to yield major discoveries.

In what follows we shall address the key issue of 
distinguishing a generic scalar-theory of gravity, as the underlying fundamental 
physical theory, from the usual general relativistic picture, where one 
expects the presence of 
fundamental scalar fields associated to inflation, dark matter and dark energy.
In order to concretely discuss the matter we
will consider a fairly general scalar-tensor theory of gravity as an example,
and indicate how its main features can be extracted from high-resolution 
measurements of the parametrized post-Newtonian (PPN) 
parameters $\beta$ and $\gamma$. As is well known, scalar-tensor 
theories of gravity mimic a plethora of 
unification models. For instance, the  graviton-dilaton system in
string/M-theory can be viewed as an specific scalar-tensor
theory of gravity. 

Of course, one should to bear in mind that current experimental data shows an impressive 
agreement with General Relativity\cite{Will05,BPT06}. 
Indeed, most stringent bounds arise from the Cassini's 2003 radiometric experiment\cite{Bertotti}: 

\beq  
\gamma - 1 = (2.1 \pm 2.5) \times 10^{-5}~~,
\label{eq:gammast3} 
\eeq    
and 

\beq  
\beta - 1 = (1.2 \pm 1.1) \times 10^{-4}
\label{eq:betast3} 
\eeq 
that arises from limits on the Strong Equivalence violation parameter, $\eta \equiv 4\beta-\gamma-3$, 
that are found to be $\eta=(4.4\pm4.5)\times 10^{-4}$, as inferred from LLR 
measurements\cite{Williams_Turyshev_Boggs_2004}.

As already mentioned, in cosmology, General Relativity allows for detailed 
predictions of the nucleosynthesis 
yields and of the properties of the CMBR, 
provided one admits the presence of fundamental 
scalar fields, the inflaton, the quintessence scalar field\cite{Copeland06} or  
the generalized Chaplygin gas model underlying scalar field, complex\cite{Bento02} or real\cite{Bertolami04a}, to account 
for the late accelerated expansion of the Universe, and 
in the case of some candidates for dark matter, self-interacting\cite{Bento01} or not\cite{Nunes99}. 
It is worth remarking the 
generalized Chaplygin gas model corresponds to a unified model of dark energy and dark matter, based on the equation of state 
$p = -A/\rho^{\alpha}$, where, $p$ is the isotropic pressure, $\rho$ is the energy density, and $A$ and $\alpha$ are 
positive phenomenological constants. Its agreement 
with observational data has been extensively studied: CMBR\cite{Bento02}, supernova\cite{Bertolami04a,Bento05}, 
gravitational lensing\cite{Silva03}, gamma-ray bursts\cite{BertSilva06} and cosmic topology\cite{Bento06}. A fully consistent 
picture for structure formation in the context of the generalized Chaplygin gas model 
remains still an open question\cite{Bento04}.
Another interesting cosmological issue concerns the resemblance of inflation and the late accelerated expansion of the Universe, 
which has lead to proposals where the inflaton 
and the quintessence scalar field are related\cite{Peebles}. 
 
A scalar field with a suitable potential can be also the way to 
explain the Pioneer anomaly\cite{Bertolami04b}. 
It is interesting to point out that scalar fields can affect 
stellar dynamics and hence, specific measurements of, for instance, 
the central temperature of stars and their luminosity can allow for 
setting bounds on scalar field models\cite{Bertolami05}.

\section{Scalar-Tensor Theories of Gravity} \label{sec:vacuum}

In many alternative theories of gravity, the gravitational
coupling strength exhibits a dependence on a field of some sort;
in scalar-tensor theories, this is a scalar field $\varphi$. 
The most general action for a scalar-tensor theory of gravity up to 
first order in the curvature can be written as

\beq  
S= {c^3\over 4\pi G}\int d^4x \sqrt{-g}
\left[\frac{1}{4}f(\varphi) R - \frac{1}{2}g(\varphi) \partial_{\mu} \varphi
\partial^{\mu} \varphi + V(\varphi) + \sum_{i}
q_{i}(\varphi)\mathcal{L}_{i}\right]~~, \label{eq:sc-tensor} 
\eeq

\noindent 
where $f(\varphi)$, $g(\varphi)$, $V(\varphi)$ are
generic functions, $q_i(\varphi)$ are coupling functions and
$\mathcal{L}_{i}$ is the Lagrangian density of the matter fields.

For simplicity, we shall consider only the theories for which 
$g(\varphi) = q_i(\varphi) = 1$. Hence, for a theory for which the $V(\varphi)$
can be locally neglected, given that its mass is fairly small so that it acts 
cosmologically, the resulting effective model can be written as 

\beq  
S= {c^3\over 4\pi G}\int d^4x \sqrt{-\hat{g}}
\left[\frac{1}{4} \hat{R} - \frac{1}{2}  \partial_{\mu} \varphi
\partial^{\mu} \varphi 
+ \sum_{i} \mathcal{L}_{i}(\hat{g}_{\mu \nu} = A^2(\varphi) g_{\mu \nu})\right]~~, 
\label{eq:sc-tensor-einsteinframe} 
\eeq 
where $A^2(\varphi)$ is the coupling function to matter and the factor that allows one to write the theory in the Einstein frame. 

It is shown that in the PPN limit, that if one writes

\beq  
ln A(\varphi) \equiv \alpha_0 (\varphi - \varphi_0) 
+ {1 \over2} \beta_0 (\varphi - \varphi_0)^2 + O(\varphi - \varphi_0)^3 ~~, 
\label{eq:mattercoupling} 
\eeq 
then\cite{Damour93}:

\beq  
\gamma -1 = -{2 \alpha_0^2 \over 1 + \alpha_0^2}~~, 
\label{eq:gammast1} 
\eeq 
and 

\beq  
\beta -1 = {1 \over 2} {\alpha_0^2 \beta_0 \over (1 + \alpha_0^2)^2}~~. 
\label{eq:betast1} 
\eeq 

Most recent bounds arising from binary pulsar PSR $B1913+16$ data indicate that\cite{Farese04}:

\beq  
\beta_0 > - 4.5~~~,~~~ \alpha_0 < 0.060 
\label{eq:gammast2} 
\eeq
and 

\beq  
{\beta -1  \over \gamma -1} < 1.1 ~~. 
\label{eq:betast2} 
\eeq 

\noindent
These results are consistent with Solar System constraints and one expects that improvement of data may allow within a 
decade to achieve $|\gamma -1| \sim 10^{-6}$, an order of magnitude better than Cassini's constraint\cite{Bertotti}.
Notice that the PPN formalism for more general cases is available\cite{Damour93}. For sure, gravitational experiments in space 
will allow to further constrain these models.

It is relevant to point out that scalar-tensor models have also been proposed to explain the 
accelerated expansion of the Universe, 
even though not quite successfully\cite{BertMartins00}.

\section{Gravitational Experiments in Space} \label{sec:exp}

Let us now give some examples of gravitational experiments that critically rely on space technology and 
that may crucially contribute to clarify some of the discussed issues. 

\subsection{Lunar Laser-Ranging: APOLLO Facility}

The Apache Point Observatory Lunar Laser-ranging Operation (APOLLO) is a
new LLR effort designed to achieve millimeter range precision and 
order-of-magnitude gains in the measurement of
physical parameters\cite{Murphy_etal_2002}.

The major advantage of APOLLO over current LLR operations
is a 3.5 m astronomical high quality telescope at a good site, the 
Sacramento Mountains of southern New Mexico (2780 m), with 
very good atmospheric quality. The APOLLO project will allow pushing
LLR into the regime of millimeter's range precision. For the Earth and
Moon orbiting the Sun, the scale of relativistic effects is set by
the ratio $(GM / r c^2)\sim v^2 /c^2 \sim 10^{-8}$.  Relativistic
effects are small compared to Newtonian effects.  The Apache Point
1 mm range accuracy corresponds to $3\times 10^{-12}$ of the
Earth-Moon distance.  The impact on gravitational
physics is expected to yield an improvement of an order of magnitude: the Equivalence
Principle would give uncertainties approaching $10^{-14}$, tests
of General Relativity effects would be smaller than $0.1$\%, and estimates of
the relative change in the gravitational constant would be about $0.1$\%
of the inverse age of the Universe. 

Therefore, the gain in the ability to conduct even more precise
tests of fundamental physics is enormous, thus this new instrument
stimulates development of better and more accurate models for the
LLR data analysis at a mm-level\cite{llr-ijmpd}.  

\subsection{The LATOR Mission}

The proposed Laser Astrometric Test Of Relativity (LATOR)\cite{solvang_lator04,Lator01,Texas@Stanford_lator,ESTEC_lator}
experiment is designed to test the metric nature of gravitation, a fundamental postulate of General
Relativity. By using a combination of independent time-series of
highly accurate gravitational deflection of light in the immediate
vicinity of the Sun, along with measurements of the Shapiro time
delay on interplanetary scales (to a precision respectively better
than $10^{-13}$ radians and 1 cm), LATOR will considerably
improve the knowledge about relativistic gravity. Its main 
objectives can be summarized as follows: i) Measure the key post-Newtonian Eddington
parameter $\gamma$ with accuracy of a part in 10$^9$, a factor
30,000 beyond the present best result, Cassini's radiometric experiment\cite{Bertotti}; 
ii) Perform the first measurement of gravity's
non-linear effects on light to about $0.01$\% accuracy; including
both the traditional Eddington $\beta$ parameter via gravity
effect on light to about $0.01$\% accuracy and also the never measured spatial
metric's second order potential contribution, $\delta$; iii) Perform a direct measurement of the solar 
quadrupole moment, $J_2$, to accuracy of a part in 200 of its
expected size; iv) Measure the ``frame-dragging''
effect on light due to the Sun's rotational gravitomagnetic field,
to $0.1$\% accuracy. LATOR's measurements will be able to push to
unprecedented accuracy the search for relevant
scalar-tensor theories of gravity by looking for a remnant scalar
field. The key element of LATOR is the
geometric redundancy provided by the laser ranging and
long-baseline optical interferometry.

LATOR mission is the 21st century version of Michelson-Morley-type 
experiment particularly suitable for the search of effects of a scalar field in
the Solar System. In spite of the previous space missions
exploiting radio waves for spacecraft tracking, this mission 
will correspond to a breakthrough in the relativistic gravity
experiments, as it allows to take full advantage of the optical
techniques that have recently became available. LATOR has a number of
advantages over techniques that use radio waves to measure
gravitational light deflection. Indeed, optical technologies allow
low bandwidth telecommunications with the LATOR spacecraft and the
use of the monochromatic light enables the observation of the
spacecraft at the limb of the Sun. The use of narrow band filters,
coronagraph optics and heterodyne detection allows for suppression of
background light to a level where the solar background is no
longer the dominant source of noise. The short wavelength allows much
more efficient links with smaller apertures, thereby eliminating
the need for a deployable antenna. Finally, the use of the International Space Station
enables the experiment to be above the Earth's atmosphere, the major source
of astrometric noise for any ground based interferometer. We think that these features fully
justify LATOR as a fundamental mission in the search for gravitational phenomena beyond General 
Relativity.

\subsection{A Mission to test the Pioneer Anomaly}

Pioneer 10 and 11 were launched in 1972 and 1973 to study the outer planets of the Solar System. 
Both probes have followed hyperbolic trajectories close to 
the ecliptic to opposite outward directions in the Solar System. 
Due to their robust design, it was possible to determine their position 
with great accuracy.
During the first years of its life, the acceleration caused by solar radiation pressure on the Pioneer 10 was 
the main effect\cite{Anderson02}. At about 20 AU (by early 1980s) solar radiation pressure became sub-dominant 
and it was possible to identify an unaccounted anomaly. 
This anomaly can be interpreted as a constant acceleration with a magnitude of 
$a=(8.74 \pm 1.33) \times 10^{-10}~m s^{-2}$ and is directed toward the Sun. 
This effect became known as the Pioneer anomaly. For the Pioneer spacecraft, it 
has been observed, at least, until $70$ AU\cite{Anderson02}. The same effect was also observed in the 
Pioneer 11 spacecraft\cite{Anderson02}.     

This puzzling deceleration has divided the space community in the last few years. If on one hand, 
skeptics have been arguing that the most likely solution for the riddle 
is some unforeseen on-board generated effect such as fuel leaking from the thrusters or non-symmetrical heat dissipation 
from the nuclear powered energy sources\cite{Scheffer}, the most optimistic point out to the fact that this 
effect may signal a new force or fundamental field of nature and hence an important window for 
new physics\footnote{The demonstration that the gravitational field due to the Kuiper Belt is not 
the cause of the anomaly has been recently reanalyzed\cite{BertVieira}. The literature is particularly rich in 
proposals\cite{Pioneer,Bertolami04b}.}. 
The approach that has been advocated by some groups that answered to the recent European Space Agency (ESA) 
call Cosmic Vision 2015 - 2025 with proposals of missions to test the Pioneer's anomalous acceleration 
is that whatever the cause of the slowing down of the spacecraft, meeting the requirements of such a 
mission would give rise to developments that will be invaluable for building and designing noise-free spacecraft 
for future deep space missions. Actually, the theoretical concept of a mission to verify the anomalous 
acceleration has been suggested earlier in a study\cite{BertTajmar02} commissioned by ESA in 2002. 

A dedicated mission would rely on a simple concept,
which consists in launching into deep space a geometrically symmetric\cite{BertTajmar02} 
and spin-stabilized\cite{Nieto-Turyshev} probe whose behavior (mechanical, thermal, electromagnetic, etc.) is carefully 
monitored. Accurate tracking of its orbit would
allow for precise evaluation of the anomaly, as any
deviation from the predicted trajectory would be used to examine
the unmodeled anomalous acceleration. The exciting possibility of using laser 
ranging techniques and the flying formation concept to characterize the nature of the 
anomaly, and solar sailing propulsion has been more recently 
discussed\cite{Pioneer}.
Particularly pleasing is the announcement that ESA is seriously considering such an 
ambitious and challenging undertaking in the period 2015 - 2025\cite{ESA}.
Naturally, a mission of this nature can be particularly useful for testing the existence of 
any Solar System range new interaction 
as well as to explore, for instance, the structure of the Kuiper Belt.

\section{Discussion and Conclusions}

Let us now review the main points of our discussion. It seems evident that resolving 
the dichotomy dark energy - dark matter versus gravity will require a concerted effort and a whole new program 
of dedicated experiments in space. 

It is an exciting prospect that dark matter can be directly detected in underground experiments or in the 
forthcoming generation of colliders. Even though it does not seem feasible to directly test the properties of 
dark energy, it is not impossible that indirect evidence can be found in laboratory. The most bold proposal suggests 
the existence of a cutoff frequency of the noise spectrum in Josephson junctions\cite{Beck}, while a more conventional 
approach is to investigate the effect that dark energy may have, for instance, on the variation of the electromagnetic 
coupling\cite{Olive}. It follows that the characterization of dark energy and dark matter will most likely 
be achieved via cosmological observations, most of them to be carried out by space-borne experiments. 
These encompass a large array of phenomena such as supernovae, gamma-ray bursts, gravitational lensing, cosmic shear, etc. 
The result of these observations will also provide increasingly detailed information on the adequacy of General Relativity 
at cosmological scales. It is quite exciting that existing supernova data\cite{Riess2004} 
together with latest CMBR data\cite{WMAP3} 
and the recently discovered baryon acoustic oscillations\cite{Eisenstein} are sufficiently constraining to virtually 
rule out\cite{Maartens}, 
for instance, most of the braneworld inspired gravity models put forward to account for the accelerated expansion of the 
Universe. The prospect of testing some of these models through the 
study of the orbital motion of planets in the Solar System has also been recently discussed\cite{Iorio}. 

We have seen how stands the situation in what concerns scalar-tensor gravity models. Relevant results are expect within a 
decade from the observation of binary pulsar systems. To further test 
General Relativity and examine the implications of its contending theories or extensions (scalar-tensor theories, 
braneworld models, string inspired models, etc.) a new program of gravity experiments in space is clearly 
needed. We have discussed how LLR can be used to improve the knowledge of relativistic gravity and 
pointed out how the LATOR mission and a mission to test the Pioneer anomaly can play a key role 
in the search for evidence of a remnant scalar field in the Solar System, to identify new forces with ranges of a 
few decades of AU and, of course, to resolve the Pioneer anomaly puzzle. 
It is relevant to point out that the latter type of mission, besides its technological appeal, 
can also to used to gather information about the vicinity of the 
Solar System as well as to set relevant upper bounds on environmental parameters such as the density of interplanetary dust and 
dark matter\cite{BertVieira}.  

\vspace{0.4cm}

\noindent
{\bf Acknowledgments}

\vspace{0.2cm}

\noindent
It is a pleasure to thank the members of the Pioneer Science Team for the countless discussions 
on the questions related with this contribution. I am particularly in debt to Jorge P\'aramos, Slava Turyshev, 
Serge Reynaud, Clovis de Matos, Pierre Toubul, Ulrich Johann and Claus L\"ammerzahl for their insights and suggestions.

\vspace{0.2cm}

\bibliographystyle{unstr}

\end{document}